\documentstyle[11pt]{article}
\begin{document}
\title{\bf UNIVERSAL FIELD EQUATIONS\\
FOR METRIC--AFFINE THEORIES OF GRAVITY}

\author{{\bf Victor Tapia}\footnote{\tt tapiens@gmail.com}\\
\\
{\it Departamento de Matem\'aticas}\\
{\it Universidad Nacional de Colombia}\\
{\it Bogot\'a, Colombia}\\
\\
and\\
\\
{\bf Maximiliano Ujevic}\footnote{\tt mujevic@ime.unicamp.br}\\
\\
{\it Instituto de Matem\'atica, Estat{\'\i}stica}\\
{\it e Computa\c c\~ao Cient{\'\i}fica}\\
{\it Universidade Estadual de Campinas}\\
{\it Campinas, S\~ao Paulo, Brazil}}

\maketitle

\begin{abstract}

We show that almost all metric--affine theories of gravity yield
Einstein equations with a non--null cosmological constant $\Lambda$.
Under certain circumstances and for any dimension, it is also possible
to incorporate a Weyl vector field $W_\mu$ and therefore the presence
of an anisotropy. The viability of these field equations is discussed
in view of recent astrophysical observations.

\end{abstract}

PACS number: 0450

\section*{}

Recent observational evidence, pointing to a non--vanishing value of
the cosmological constant \cite{OS95} and to the existence of a
preferred direction \cite{NR97} in the Universe, has reopened the
debate on which is the `most correct' theory of gravitation. As in any
field theory there are two aspects to be considered: the field
equations and the variational formulation of them. Undoubtedly, general
relativity is the best theory we have for the description of
gravitational phenomena. The Einstein field equations were obtained on
a phenomenological basis and therefore they exhibit a strong
observational agreement. On the other hand, the variational formulation
of Einstein field equations provides a systematic way to deal with
them. However, this formulation must not be considered as fundamental
as the field equations themselves and, in fact, several of the problems
of general relativity come from its variational formulation. Just to
mention a well known problem, the `unphysical' non--renormalizability
of general relativity is obtained from a dimensional analysis of the
Einstein--Hilbert Lagrangian even when the Einstein field equations
describe `physically' sensible situations. Therefore, in order to
accommodate the new observational data, not describable in the context
of general relativity, it would be better to start by looking for field
equations with a better observational agreement than just on the basis
of theoretical considerations. Of course, many {\it ad hoc} field
equations can be conceived to agree with observation and in this case a
phenomenological argument, such as that leading from Newtonian gravity
to general relativity, is lacking. In the face of this situation, and
in spite of the comments above, we must allow our search for the needed
modifications to be guided with the help of theoretical considerations.

It has been shown recently \cite{BFFV} that several families of
gravitational Lagrangians give rise to the same field equations, namely
Einstein field equations with a non--null cosmological constant
$\Lambda$. Therefore, these field equations are better suited to
accommodate the recent observations concerning a non--null value of
$\Lambda$. This `universality' property was originally observed for
Lagrangians depending in an arbitrary way from
$\bigl<R\bigr>=g^{\mu\nu}R_{\mu\nu}({\bf\Gamma})$
\cite{Li83,HB93,FFV1,FFV2}. Later on this property was established for
a wider family of Lagrangians depending at most on 
$\bigl<R^2\bigr>=R_{\mu\nu}({\bf\Gamma})R^{\mu\nu}({\bf\Gamma})$
\cite{BFFV}.

In the present work, we show that these universal field equations can
be obtained from a family of variational principles which is even wider
than that considered previously. In addition, almost any metric--affine
Lagrangian will serve the prupose of providing a variational
description of them. Furthermore, we show that, under certain
circumstances, it is also possible to incorporate an anisotropy through
the appearance of a Weyl vector related to conformal invariance. In
previous works this possibility was established only for $n=2$, where
$n$ is the dimension of the base space. Here we show that this can be
done in any dimension. This result is particularly relevant for $n=4$
since a Weyl vector may describe the recently observed anisotropy of
the Universe.

The freedom in the choice of the Lagrangian provides us with a
theoretical setting to incorporate almost any other desirable property
of a viable gravitational theory, in particular renormalizability and
scale invariance, without restricting the field equations.

Let us start with a brief reminder of the state of the art in canonical
gravity. A large family of different situations in gravitational
physics is correctly described by the Einstein field equations. The
problem of obtaining Einstein field equations from a variational
principle was first addressed, almost contemporarily, by Einstein
\cite{Ei15} and by Hilbert \cite{Hi15}; cf. \cite{FFR} for an
interesting historical account. Under a suitable set of simplifying
assumptions they concluded that a possible Lagrangian for the Einstein
field equations is given by

\begin{equation}
{\cal L}_{EH}=R({\bf g})\,g^{1/2}\,,
\label{01}
\end{equation}

\noindent where $R({\bf g})=g^{\mu\nu}R_{\mu\nu}({\bf g})$ is the Ricci
scalar constructed from  a purely metric Ricci tensor $R_{\mu\nu}({\bf
g})$. Almost inmediately, Palatini \cite{Pa19} presented an alternative
Lagrangian yielding the same field equations, namely

\begin{equation}
{\cal L}_P=g^{\mu\nu}\,R_{\mu\nu}({\bf\Gamma})\,g^{1/2}\,,
\label{02}
\end{equation}

\noindent where $R_{\mu\nu}({\bf\Gamma})$ is the symmetric part of the
Ricci tensor for a connection ${\Gamma^\lambda}_{\mu\nu}$ symmetric in
its two lower indices.

For later convenience we will display our convention for the Ricci
tensor. The starting point is the Riemann tensor given by

\begin{equation}
{R^\lambda}_{\rho\mu\nu}({\bf\Gamma})=\partial_\mu{\Gamma^\lambda}_{\nu\rho}
-\partial_\nu{\Gamma^\lambda}_{\mu\rho}+{\Gamma^\lambda}_{\mu\sigma}\,{\Gamma^\sigma}_{\nu\rho}
-{\Gamma^\lambda}_{\nu\sigma}\,{\Gamma^\sigma}_{\mu\rho}\,, \label{03}
\end{equation}

\noindent and its is obviously antisymmetric in its last two indices.
From contractions of the Riemann tensor alone, not involving any
auxiliary field, two second--rank tensors can be obtained, namely
$F_{\mu\nu}={R^\lambda}_{\lambda\mu\nu}=\partial_\mu\Gamma_\nu-\partial_\nu\Gamma_\mu$,
with $\Gamma_\mu={\Gamma^\lambda}_{\mu\lambda}$, and the Ricci tensor

\begin{equation}
R_{\mu\nu}({\bf\Gamma})={R^\lambda}_{\mu\lambda\nu}({\bf\Gamma})=\partial_\lambda{\Gamma^
\lambda}_{\nu\mu}-\partial_\nu{\Gamma^\lambda}_{\lambda\mu}+{\Gamma^\lambda}_{\lambda\sigma}
\,{\Gamma^\sigma}_{\nu\mu}-{\Gamma^\lambda}_{\nu\sigma}\,{\Gamma^\sigma}_{\lambda\mu}\,.
\label{04}
\end{equation}

\noindent In the rest of the work we will restrict our considerations
to the symmetric part of the Ricci tensor (\ref{04}) for a connection
${\Gamma^\lambda}_{\mu\nu}$ symmetric in its two lower indices.

The Lagrangians (\ref{01}) and (\ref{02}) both lead to the same set of
field equations, namely, the Einstein field equations

\begin{equation}
R_{\mu\nu}({\bf g})-{1\over2}\,R({\bf g})\,g_{\mu\nu}=0\,.
\label{05}
\end{equation}

\noindent After considering the trace of equation (\ref{05}), which
implies $R({\bf g})=0$, the Einstein field equations can be rewritten
in the simplified, but equivalent form

\begin{equation}
R_{\mu\nu}({\bf g})=0\,.
\label{06}
\end{equation}

\noindent The exception is the dimension $n=2$ in which case (\ref{05})
is an identity. Equations (\ref{06}) have a strong phenomenological
support since, in the non--relativistic limit, they are equivalent to
Newtonian gravity and furthermore they have been tested and found to be
correct in a variety of situations beyond the realm of Newtonian
gravity \cite{Wi92}. However, due to several reasons stemming from the
need to obtain better observational agreements \cite{AF92} to formal
problems \cite{DW67}, such as those found in quantum gravity,
alternative theories need to be seriously considered.

The first modification of generqal relativity is due to Einstein
himself \cite{Ei17} and involves the introduction of a cosmological
constant $\Lambda$. In this case, the Lagrangians (\ref{01}) and
(\ref{02}) are modified to

\begin{equation}
{\cal L}_{EH\Lambda}=[R({\bf g})-\Lambda_n]\,g^{1/2}\,,
\label{07}
\end{equation}

\noindent and

\begin{equation}
{\cal L}_{P\Lambda}=[g^{\mu\nu}\,R_{\mu\nu}({\bf\Gamma})-\Lambda_n]\,g^{1/2}\,,
\label{08}
\end{equation}

\noindent where $n$, as in the rest of this work, denotes the dimension
of the base space. In both cases the modified field equations are

\begin{equation}
R_{\mu\nu}({\bf g})-{1\over2}\,[R({\bf g})-\Lambda_n]\,g_{\mu\nu}=0\,.
\label{09}
\end{equation}

\noindent Considering the trace of this equation, which implies $R({\bf
g})=n\Lambda_n/(n-2)$, the Einstein field equations, with a
cosmological constant, can be rewritten in the simplified form

\begin{equation}
R_{\mu\nu}({\bf g})-{1\over{(n-2)}}\,\Lambda_n\,g_{\mu\nu}=0\,.
\label{10}
\end{equation}

\noindent The exception is the case for $n=2$ in which the first two
terms in (\ref{09}) are an identity and therefore $\Lambda_2=0$. In
view of this result it is more convenient to write
$\Lambda_n=(n-2)\Lambda$. With this new parametrization the field
equations (\ref{10}) are rewritten as

\begin{equation}
R_{\mu\nu}({\bf g})-\Lambda\,g_{\mu\nu}=0\,,
\label{11}
\end{equation}

\noindent and the trace is given simply by $R({\bf g})=n\Lambda$. Our
later generalizations will be based on this simplified, but equivalent,
form of Einstein field equations.

As mentioned above, recent observations \cite{OS95} point to a
non--null value of the cosmological constant $\Lambda$ and therefore
(\ref{11}), rather than (\ref{06}), need to be considered as the
correct field equations for the gravitational field. A small, but
non--null, value of the cosmological constant is most welcome since it
is able to solve several observational problems, such as conflicting
ages of stars compared with the age of the Universe \cite{Fr94}. The
Newtonian limit remains unmodified since $\Lambda$ is small enough to
play a significant role at that regime.

The Lagrangians (\ref{07}) and (\ref{08}) are correct classically,
since they lead to the correct field equations. However, due to several
reasons, mainly non--renormalizability, they do not serve as a starting
point for a quantization program. As remarked by Martellini
\cite{Ma83}, what is really relevant for a variational formulation of a
gravitational theory is to obtain the correct classical equations, in
this case (\ref{11}), but the Lagrangian is completely unrestricted for
the rest and may even strongly differ from the simple prescriptions
contained in (\ref{07}) and (\ref{08}). This remark is most important
when one looks for better starting points for a quantum gravity
programme.

The simplest model pointing to satisfy the renormalizability
requirements necessary for quantum gravity is the Lagrangian $R^2$. In
the purely metric case we have

\begin{equation}
{\cal L}_{R^2({\bf g})}=[\alpha\,R^2({\bf g})+\beta\,R_{\mu\nu}({\bf g})\,R^{\mu\nu}({\bf g})
+\gamma\,R_{\mu\nu\lambda\rho}({\bf g})\,R^{\mu\nu\lambda\rho}({\bf g})]\,g^{1/2}\,.
\label{12}
\end{equation}

\noindent This Lagrangian has been extensively considered from the
point of view of necessary counterterms for achieving the
renormalizability of the Einstein--Hilbert Lagrangian. However, the
Lagrangian $R^2$ can be considered by itself, without any reference to
general relativity, avoiding in this way the introduction of
dimensionful coupling constants \cite{UD62}. In this case it is
possible to establish an equivalence between (\ref{12}) and general
relativity coupled to some kind of matter. However, the Lagrangian
(\ref{12}) is non--polynomial in the metric field and this makes it
useless for a quantization programme \cite{Ma83}.

The situation is much better for a metric--affine theory where the
starting point is the Lagrangian

\begin{equation}
{\cal L}_{R^2({\bf\Gamma})}=[\alpha\,\bigl<R\bigr>^2+\beta\,\bigl<R^2\bigr>+\gamma\,{\rm Rie}^2]\,g^{1/2}\,;
\label{13}
\end{equation}

\noindent $\bigl<R\bigr>=g^{\mu\nu}R_{\mu\nu}({\bf\Gamma})$, 
$\bigl<R^2\bigr>=R_{\mu\nu}({\bf\Gamma})R^{\mu\nu}({\bf\Gamma})$, and ${\rm Rie}^2$ denotes all possible, algebraically independent, contractions of the Riemann tensor ${R^\lambda}_{\rho\mu\nu}({\bf\Gamma})$ and since it has fewer symmetries than the Riemann--Christoffel tensor $R_{\lambda\rho\mu\nu}({\bf g})$, more terms of that kind need to be considered in (\ref{13}). In this case it is possible to establish the equivalence between (\ref{13}) and general relativity with a cosmological constant, as we will shown below. The Lagrangian (\ref{13}) is far more appealing than (\ref{12}) since it is polynomial in ${\bf\Gamma}$ and therefore more amenable for a quantum treatment \cite{Ma83}. The terms ${\rm Rie}^2$ are Yang--Mills--like and they can be dealt with in that context, see \cite{HMMN}. The terms containing $\bigl<R\bigr>$ and $\bigl<R^2\bigr>$ only, were first considered by Stephenson \cite{St58} and by Higgs \cite{Hi59}.

Lagrangians with a more general dependence on $\bigl<R\bigr>$ and
$\bigl<R^2\bigr>$ have appeared in several studies related to the
gravitational field equations. A generic Lagrangian of the form ${\cal
L}_{\bigl<R\bigr>}=f(\bigl<R\bigr>)g^{1/2}$ was considered by Lim
\cite{Li83} and, under a simple assumption, he obtained Einstein field
equations with a cosmological constant, namely (\ref{11}). This was a
first indication that one can consider a wider family of variational
principles from which to obtain gravitational equations. For a
restricted family of Lagrangians ${\cal L}_{\bigl<R\bigr>}$, Hamity and
Barraco \cite{HB93} obtained similar field equations in which, in
addition to the cosmological constant $\Lambda$, conformal invariance
was also incorporated. Finally, Ferraris {\it et al.} \cite{FFV1,FFV2}
has established the important result that all Lagrangians ${\cal
L}_{\bigl<R\bigr>}$ possess `universal' field equations corresponding
to Einstein field equations with a cosmological constant $\Lambda$,
namely (\ref{11}). Recently, the universality of the field equations
(\ref{11}) was also proved for Lagrangians with an arbitary dependence
on $\bigl<R^2\bigr>$ \cite{BFFV}.

These results could be interpreted as an indication that the
cosmological constant and other desirable properties of the
gravitational theory may be obtained as properties of the field
equations, rather than being incorporated into the Lagrangian.

In dimension $n=2$, it is possible to incorporate a Weyl vector $W_\mu$
related to conformal invariance. The presence of a vector field
indicates the existence of a preferred direction. For $n=4$ a vector
field would be a most welcome ingredient in the field equations since
this might describe the recently observed anisotropy of the universe
\cite{NR97}.

In the present work we show that the universal field equations
(\ref{11}) can be obtained from a wider family of Lagrangians given by
${\cal L}=L(\bigl<{\bf R}\bigr>)g^{1/2}$, with an arbitrary dependence
on ${\bigl<R\bigr>^\mu}_\nu=g^{\mu\lambda}R_{\lambda\nu}({\bf\Gamma})$.
Therefore, it may depend on $\bigl<R\bigr>$ and $\bigl<R^2\bigr>$, but
also on higher--order powers of  ${\bigl<R\bigr>^\mu}_\nu$. More
importantly, we show that for any dimension, in particular for $n=4$,
under certain circumstances, it is possible to incorporate a Weyl
vector field $W_\mu$. This is also possible in the situations studied
previously by other authors, however, they did not mention this
possibility. The Weyl vector field $W_\mu$ contributes new terms to the
field equations which finally read

\begin{eqnarray}
R_{\mu\nu}({\bf g})-{1\over4}\,(n-2)\,(\nabla_\mu W_\nu+\nabla_\nu W_\mu)+{1\over4}\,(n-2)\,W_\mu\,W_\nu&&\nonumber\\
-\left({1\over2}\,(\nabla W)+{1\over4}\,(n-2)\,W^2\right)\,g_{\mu\nu}-\Lambda\,g_{\mu\nu}&=&0\,,
\label{14}
\end{eqnarray}

\noindent where $(\nabla W)=g^{\mu\nu}\nabla_\mu W_\nu$ and
$W^2=g^{\mu\nu}W_\mu W_\nu$. For the special case $n=4$ the field
equations above reduce to

\begin{equation}
R_{\mu\nu}({\bf g})-{1\over2}\,(\nabla_\mu W_\nu+\nabla_\nu W_\mu)+{1\over2}\,W_\mu\,W_\nu-{1\over2}\,\left((\nabla W)+W^2\right)\,g_{\mu\nu}-\Lambda\,g_{\mu\nu}=0\,.
\label{15}
\end{equation}

\noindent Our method of proof is essentially different from the one
used in \cite{BFFV,FFV1,FFV2} and is strongly based on a systematic use
of the Cayley--Hamilton theorem, which we will now describe.

Let us review some fundamental results related to the Cayley--Hamilton
theorem. Let us start by considering a generic $n\times n$ matrix
${\bigl<R\bigr>^\mu}_\nu$. Higher powers of this matrix are denoted by

\begin{equation}
{\bigl<R^k\bigr>^\mu}_\nu=\underbrace{{\bigl<R\bigr>^\mu}_\circ\,\cdots\,
{\bigl<R\bigr>^\circ}_\nu}_{k\,{\rm times}}\,.
\label{16}
\end{equation}

\noindent Accordingly, we denote the traces by

\begin{equation}
\bigl<R^k\bigr>={\bigl<R\bigr>^\mu}_\mu\,.
\label{17}
\end{equation}

\noindent Let us introduce next the cumulant coefficients
$\bigl<C^k\bigr>$ given by

\begin{eqnarray}
\bigl<C^0\bigr>&=&1\,,\nonumber\\
\bigl<C^1\bigr>&=&\bigl<R\bigr>\,,\nonumber\\
\bigl<C^2\bigr>&=&{1\over2}\,\left(\bigl<R\bigr>^2-\bigl<R^2\bigr>\right)\,,\nonumber\\
\bigl<C^3\bigr>&=&{1\over{3!}}\,\left(\bigl<R\bigr>^3-3\,\bigl<R\bigr>\,\bigl<R^2\bigr>
+2\,\bigl<R^3\bigr>\right)\,,\nonumber\\
\bigl<C^4\bigr>&=&{1\over{3!}}\,\left(\bigl<R\bigr>^4-6\,\bigl<R\bigr>^2\,\bigl<R^2\bigr>
+8\,\bigl<R\bigr>\,\bigl<R^3\bigr>+3\,\bigl<R^2\bigr>^2-6\,\bigl<R^4\bigr>\right)\,,\nonumber\\
&\vdots&\,.
\label{18}
\end{eqnarray}

\noindent Essentially the cumulants relate the coefficients in a Taylor
series with those of a Fourier expansion. some of the cumulants
(\ref{18}) are familiar objects in matrix calculus; namely,
$\bigl<C^1\bigr>={\rm tr}\left({\bigl<R\bigr>^\mu}_\nu\right)$ and 
$\bigl<C^n\bigr>=\det\left({\bigl<R\bigr>^\mu}_\nu\right)$. A further
property of the cumulants is that, for a given $n$, only the first $n$
cumulants are non--trivial, while upper cumulants are identically zero,
$\bigl<C^k\bigr>\equiv0$, $k>n$.

The Cayley--Hamilton theorem \cite{Ne70} states that only the first
$(n-1)$ powers of ${\bigl<R\bigr>^\mu}_\nu$ are linearly independent.
The relation among them is the characteristic polynomial

\begin{equation}
\sum_{k=0}^n\,(-)^k\,\bigl<C^k\bigl>\,{\bigl<R^{n-k}\bigr>^\mu}_\nu\equiv0\,.
\label{19}
\end{equation}

\noindent Accordingly, only the first $n$ traces (\ref{17}) are
algebraically independent.

For the purposes of this work we introduce a further set of  relevant
quantities which are obtained by considering derivatives of the
cumulants $\bigl<C^k\bigr>$ as follows. Let

\begin{equation}
{\bigl<C^k\bigl>^\mu}_\nu=\left({{\partial\bigl<C^k\bigr>}\over{\partial(\delta^\nu_\mu)}}
\right)=g^{\mu\lambda}\,{{\partial\bigl<C^k\bigr>}\over{\partial g^{\lambda\nu}}}\,.
\label{20}
\end{equation}

\noindent The first $\bigl<C^k\bigr>$'s are given by

\begin{eqnarray}
{\bigl<C^1\bigr>^\mu}_\nu&=&{\bigl<R\bigr>^\mu}_\nu\,,\nonumber\\
{\bigl<C^2\bigr>^\mu}_\nu&=&\bigl<R\bigr>\,{\bigl<R\bigr>^\mu}_\nu-{\bigl<R^2\bigr>^\mu}_\nu\,,
\nonumber\\
&=&\bigl<C^1\bigr>\,{\bigl<C^1\bigr>^\mu}_\nu-{\bigl<C^1\bigr>^\mu}_\lambda\,
{\bigl<C^1\bigr>^\lambda}_\nu\,,\nonumber\\
{\bigl<C^3\bigr>^\mu}_\nu&=&{1\over2}\,\left(\bigl<R\bigr>^2-\bigl<R^2\bigr>\right)
\,{\bigl<R\bigr>^\mu}_\nu+\left({\bigl<R^2\bigr>^\mu}_\lambda-
\bigl<R\bigr>\,{\bigl<R\bigr>^\mu}_\lambda\right)\,{\bigl<R\bigr>^\lambda}_\nu\nonumber\\
&=&\bigl<C^2\bigr>\,{\bigl<C^1\bigr>^\mu}_\nu-{\bigl<C^2\bigr>^\mu}_\lambda\,
{\bigl<C^1\bigr>^\lambda}_\nu\,,\nonumber\\
&\vdots&\,.
\label{21}
\end{eqnarray}

\noindent The equation

\begin{equation}
{\bigl<C^n\bigr>^\mu}_\nu=\bigl<C^n\bigr>\,\delta^\mu_\nu\,,
\label{22}
\end{equation}

\noindent is particularly interesting and is a statement similar to
(\ref{19}) but in terms of cumulants. The product of two
$\bigl<C\bigr>$'s follows the rule

\begin{equation}
{\bigl<C^i\bigr>^\mu}_\lambda\,{\bigl<C^j\bigr>^\lambda}_\nu=\sum_{\ell=0}^{k-1}\,
\left[\bigl<C^{i+j-(\ell+1)}\bigr>\,{\bigl<C^{\ell+1}\bigr>^\mu}_\nu
-{\bigl<C^\ell\bigr>^\mu}_\lambda\,{\bigl<C^{i+j-\ell}\bigr>^\lambda}_\nu\right]\,,
\label{23}
\end{equation}

\noindent where $\ell={\rm min}(i,j)$. Another interesting relation is
given by

\begin{equation}
{{\partial\bigl<C^k\bigr>}\over{\partial R_{\mu\nu}}}=\bigl<C^{k-1}\bigr>\,g^{\mu\nu}-
\bigl<C^{k-1}\bigr>^{\mu\nu}\,.
\label{24}
\end{equation}

The relations (\ref{21}) can be inverted to express $\bigl<R\bigr>$'s
in terms of $\bigl<C\bigr>$'s. Relations (\ref{18}) are inverted to

\begin{eqnarray}
\bigl<R\bigr>&=&\bigl<C^1\bigr>\,,\nonumber\\
\bigl<R^2\bigr>&=&\bigl<C^1\bigr>^2-2\,\bigl<C^2\bigr>\,,\nonumber\\
\bigl<R^3\bigr>&=&\bigl<C^1\bigr>^3-3\,\bigl<C^1\bigr>\,\bigl<C^2\bigr>+3\,\bigl<C^3\bigr>\,,
\nonumber\\
&\vdots&\,,
\label{25}
\end{eqnarray}

\noindent while relations (\ref{21}) to

\begin{eqnarray}
{\bigl<R\bigr>^\mu}_\nu&=&{\bigl<C^1\bigr>^\mu}_\nu\,,\nonumber\\
{\bigl<R^2\bigr>^\mu}_\nu&=&\bigl<C^1\bigr>\,{\bigl<C^1\bigr>^\mu}_\nu-{\bigl<C^2\bigr>^\mu}_\nu
\,,\nonumber\\
{\bigl<R^3\bigr>^\mu}_\nu&=&\left(\bigl<C^1\bigr>^2-\bigl<C^2\bigr>\right)\,{\bigl<C^1\bigr>^
\mu}_\nu-\bigl<C^1\bigr>\,{\bigl<C^2\bigr>^\mu}_\nu+{\bigl<C^3\bigr>^\mu}_\nu\,,\nonumber\\
&\vdots&\,.
\label{26}
\end{eqnarray}

\noindent Finally we can write the characteristic polynomial (\ref{19})
in terms of $\bigl<C\bigr>$'s just to obtain (\ref{22}), showing that
this is, in fact, a statement equivalent to (\ref{19}) in terms of the
cumulants $\bigl<C\bigr>$'s.

In view of the above results we are now ready to write the most general
Lagrangian with an arbitrary dependence on ${\bigl<R\bigr>^\mu}_\nu$.
We must consider Lagrangians with a dependence restricted only to the
first $n$ traces $\bigl<R^k\bigr>$, $k=1,\cdots,n$ and this is
equivalently done in terms of the $\bigl<C^k\bigr>$'s, $k=1,\cdots,n$.
Namely

\begin{equation}
{\cal L}_n=L_n\left(\bigl<C^1\bigr>,\,\cdots,\,\bigl<C^n\bigr>\right)\,g^{1/2}\,.
\label{27}
\end{equation}

\noindent Now we must consider a Palatini--like variational principle
in which the metric and the connection are varied independently.

Variation with respect to the metric gives

\begin{equation}
{\bigl<C^{n-1}\bigr>^\mu}_\nu\,{{\partial L_n}\over{\partial\bigl<C^{n-1}\bigr>}}+\cdots+
{\bigl<C^1\bigr>^\mu}_\nu\,{{\partial L_n}\over{\partial\bigl<C^1\bigr>}}+\left(
\bigl<C^n\bigr>\,{{\partial L_n}\over{\partial\bigl<C^n\bigr>}}-{1\over2}\,L_n\right)\,\delta^\mu_\nu=0\,,
\label{28}
\end{equation}

\noindent where we have used equation (\ref{22}). In order to
completely determine the tensor ${\bigl<R\bigr>^\mu}_\nu$ we need to
know its first $(n-1)$ powers and for this purpose it is necessary to
have $(n-2)$ other relations similar to (\ref{28}). We may obtain them
by contracting equation (\ref{28}) with ${\bigl<C^k\bigr>^\mu}_\nu$,
$k=1,\cdots,n-2$ and using (\ref{22}) and (\ref{23}) to reduce higher
powers of $C$'s. We obtain a set of $n$ relations of the form

\begin{equation}
\left(\matrix{{D^1}_1&\cdots&{D^1}_{n-1}\cr\vdots&&\vdots\cr
{D^{n-1}}_1&\cdots&{D^{n-1}}_{n-1}\cr}\right)\,
\left(\matrix{{\bigl<C^1\bigr>^\mu}_\nu\cr\vdots\cr
{\bigl<C^{n-1}\bigr>^\mu}_\nu\cr}\right)=
\left(\matrix{E^1\cr\vdots\cr E^{n-1}\cr}\right)\,\delta^\mu_\nu\,,
\label{29}
\end{equation}

\noindent where the coefficients $D$'s and $E$'s are functions of
$\bigl<C\bigr>$'s.

Some exceptional cases may appear. The first one is given by the
condition $\det(D)=0$, but this is satisfied only for a restricted
family of Lagrangians with null measure in the space of functions.
Another exceptional situation is obtained when equation (\ref{28}) is
an identity. In this case, no algebraic restrictions are imposed over
the Ricci tensor. The solution is

\begin{equation}
{\cal L}_E=\sqrt{\bigm<C^n\bigr>}\,g^{1/2}=\sqrt{\det[R_{\mu\nu}({\bf\Gamma})]}\,.
\label{30}
\end{equation}

\noindent This is the Eddington Lagrangian \cite{Ed23} which does not
depend on the metric, in agreement with the fact that (\ref{28}) is
satisfied identically.

The regular case is characterized by $\det(D)\not=0$ and we arrive at
the relations

\begin{equation}
{\bigl<C^k\bigr>^\mu}_\nu=\bigl<F^k\bigr>\,\delta^\mu_\nu\,,
\label{31}
\end{equation}

\noindent with $\bigl<F\bigr>$'s functions of $\bigl<C\bigr>$'s.
Therefore, under quite general assumptions the field equations
(\ref{28}) are equivalent to (\ref{31}). However, this result can be
simplified even more since for $k=1$, equation (\ref{31}) reduces to

\begin{equation}
{\bigl<C^1\bigr>^\mu}_\nu={\bigl<R\bigr>^\mu}_\nu=\bigl<F^1\bigr>\,\delta^\mu_\nu\,.
\label{32}
\end{equation}

\noindent The cumulants (\ref{18}) then reduce to

\begin{equation}
\bigl<C^k\bigl>={{n!}\over{(n-k)!k!}}\,\bigl<F^1\bigr>^k={{n!}\over{k!(n-k)!}}\,\left[{1\over n}
\,\bigl<R\bigr>\right]^k\,.
\label{33}
\end{equation}

\noindent Therefore, equation (\ref{32}) is rewritten as

\begin{equation}
R_{\mu\nu}({\bf\Gamma})=F\left(\bigl<R\bigr>\right)\,g_{\mu\nu}\,,
\label{34}
\end{equation}

\noindent where $F\left(\bigl<R\bigr>\right)$ is some function of
$\bigl<R\bigr>$. The trace of this equation gives $\bigl<R\bigr>=n
F\left(\bigl<R\bigr>\right)$. Two important situations may appear:

\begin{enumerate}

\item If $\bigl<R\bigl>={\rm constant}$, then equation (\ref{34})
reduces to

\begin{equation}
R_{\mu\nu}({\bf\Gamma})-\Lambda\,g_{\mu\nu}=0\,,
\label{35}
\end{equation}

with $\Lambda={\rm constant}$. Equation (\ref{35}) above looks the same
as the Einstein field equations (\ref{11}). However, they are not yet
equivalent since a metricity condition, {\it i.e.}, a relationship
between ${\bf\Gamma}$ and ${\bf g}$, is still required. This
relationship will be obtained starting from equation (\ref{46}).

\item The second situation appears when the trace of equation
(\ref{34}) is identically zero. This means that (\ref{34}) must be of
the form

\begin{equation}
R_{\mu\nu}({\bf\Gamma})-{1\over n}\,\bigl<R\bigr>\,g_{\mu\nu}=0\,.
\label{36}
\end{equation}

This last situation is the most important for physical applications and
we will return to it below.

\end{enumerate}

As a way to verify the correctness of the procedure developed above,
let us consider in detail the case $n=3$. The Lagrangian is given by

\begin{equation}
{\cal L}_3=L_3\left(\bigl<C^1\bigr>,\,\bigl<C^2\bigr>,\,\bigl<C^3\bigr>\right)\,g^{1/2}\,.
\label{37}
\end{equation}

\noindent Variation with respect to the metric gives

\begin{equation}
{\bigl<C^1\bigr>^\mu}_\nu\,{{\partial L_3}\over{\partial\bigl<C^1\bigr>}}
+{\bigl<C^2\bigr>^\mu}_\nu\,{{\partial L_3}\over{\partial\bigl<C^2\bigr>}}
+\left(\bigl<C^3\bigr>\,{{\partial L_3}\over{\partial\bigl<C^3\bigr>}}-{1\over2}\,L_3\right)\,\delta^\mu_\nu=0\,,
\label{38}
\end{equation}

\noindent where we have used (\ref{22}) for $n=3$. The trace of this
equation gives

\begin{equation}
\bigl<C^3\bigr>\,{{\partial L_3}\over{\partial\bigl<C^3\bigr>}}-{1\over2}\,L_3=-{1\over3}\,\left[2\,\bigl<C^2\bigr>\,{{\partial L_3}\over{\partial\bigl<C^2\bigr>}}
+\bigl<C^1\bigr>\,{{\partial L_3}\over{\partial\bigl<C^1\bigr>}}\right]\,.
\label{39}
\end{equation}

\noindent Therefore, equation (\ref{38}) is rewritten as

\begin{equation}
{\bigl<C^1\bigr>^\mu}_\nu\,{{\partial L_3}\over{\partial\bigl<C^1\bigr>}}
+{\bigl<C^2\bigr>^\mu}_\nu\,{{\partial L_3}\over{\partial\bigl<C^2\bigr>}}
-{1\over3}\,\left[2\,\bigl<C^2\bigr>\,{{\partial L_3}\over{\partial\bigl<C^2\bigr>}}
+\bigl<C^1\bigr>\,{{\partial L_3}\over{\partial\bigl<C^1\bigr>}}\right]\,\delta^\mu_\nu=0\,.
\label{40}
\end{equation}

\noindent Let us now contract this equation with
${\bigl<C^1\bigr>^\mu}_\nu$. Using (\ref{23}) we then obtain

\begin{equation}
{1\over3}\,\left[\bigl<C^2\bigr>\,{{\partial L_3}\over{\partial\bigl<C^2\bigr>}}+2\,\bigl<
C^1\bigr>\,{{\partial L_3}\over{\partial\bigl<C^1\bigr>}}\right]\,{\bigl<C^1\bigr>^\mu}_\nu
-{{\partial L_3}\over{\partial\bigl<C^1\bigr>}}\,{\bigl<C^2\bigr>^\mu}_\nu
-\bigl<C^3\bigr>\,{{\partial L_3}\over{\partial\bigl<C^2\bigr>}}\,\delta^\mu_\nu=0\,.
\label{41}
\end{equation}

\noindent Equations (\ref{40}) and (\ref{41}) allow to solve for
${\bigl<C^1\bigr>^\mu}_\nu$ and ${\bigl<C^2\bigr>^\mu}_\nu$, as in
(\ref{31}), if

\begin{equation}
\Delta={1\over3}\,{{\partial L_3}\over{\partial\bigl<C^2\bigr>}}\,\left[\bigl<C^2\bigr>\,{{\partial L_3}\over{\partial
\bigl<C^2\bigr>}}+2\,\bigl<C^1\bigr>\,{{\partial L_3}\over{\partial\bigl<C^1\bigr>}}\right]
+\left({{\partial L_3}\over{\partial\bigl<C^1\bigr>}}\right)^2\not=0\,.
\label{42}
\end{equation}

\noindent The condition $\Delta=0$ is fulfilled only by a restricted
family of Lagrangians, namely,

\begin{equation}
L_3=\bigl<C^1\bigr>\,u^{1/3}\,(4\,u-1)^{1/6}\,\left({{1+\sqrt{1-3u}}\over{1-\sqrt{1-3u}}}
\right)^{\pm1/3}\,\left({{1+2\sqrt{1-3u}}\over{1-2\sqrt{1-3u}}}\right)^{\mp1/6}\,G\left(\bigl<
C^3\bigr>\right)\,.
\label{43}
\end{equation}

\noindent where $u=\bigl<C^2\bigr>/\bigl<C^1\bigr>^2$. Only one sign 
can be chosen in (\ref{43}) and $G$ is determined from equation
(\ref{39}). This shows that our derivation is correct, except for a
null--measure set.

Let us now turn to the second set of field equations, the metricity
condition, which must be obtained from (\ref{27}). Variation with
respect to the connection ${\bf\Gamma}$ gives

\begin{equation}
\nabla_\lambda\left(\gamma^{\mu\nu}\,g^{1/2}\right)=0\,,
\label{44}
\end{equation}

\noindent with

\begin{eqnarray}
\gamma^{\mu\nu}&=&{{\partial L_n}\over{\partial R_{\mu\nu}}}\nonumber\\
&=&g^{\mu\nu}\,{{\partial L_n}\over{\partial\bigl<C^1\bigr>}}
+\left(\bigl<C^1\bigr>\,g^{\mu\nu}-\bigl<C^1\bigr>{\mu\nu}\right)\,{{\partial L_n}\over{\partial\bigl<C^2\bigr>}}+\cdots+
+\left(\bigl<C^{n-1}\bigr>\,g^{\mu\nu}-\bigl<C^{n-1}\bigr>{\mu\nu}\right)\,{{\partial L_n}\over{\partial\bigl<C^n\bigr>}}\,,\nonumber\\
&&
\label{45}
\end{eqnarray}

\noindent where we have used (\ref{24}). This is the Eddington
prescription \cite{Ed23} to define a metric in a metric--affine, or
purely affine, theory. Let us now consider the two different situations
which appeared from (\ref{34}).

\begin{enumerate}

\item Combining equation (\ref{45}) with equation (\ref{35}) we obtain
$\gamma_{\mu\nu}=K g_{\mu\nu}$, where $K={\rm constant}$. Therefore,
$g_{\mu\nu}$ satisfies equation (\ref{44}), with
$\gamma_{\mu\nu}=g_{\mu\nu}$. The solution to (\ref{44}) is

\begin{equation}
{\Gamma^\lambda}_{\mu\nu}=\left\{{}^{\,\lambda}_{\mu\nu}\right\}({\bf g})={1\over2}\,g^{\lambda
\rho}\,\left(\partial_\mu g_{\nu\rho}+\partial_\nu g_{\mu\rho}-\partial_\rho g_{\mu\nu}\right)
\label{46}
\end{equation}

and corresponds to the Christoffel symbol for the metric tensor
$g_{\mu\nu}$. Replacing (\ref{46}) into (\ref{35}) we obtain the
universal field equations (\ref{11}). The exception is $n=2$ and this
case can be considered under the following point.

\item The second case is much more interesting. In this case the
relevant equation is (\ref{36}). This equation implies that there is a
fundamental invariance in the theory. In fact, nothing changes under
the transformation $g_{\mu\nu}\rightarrow\Omega({\bf x})g_{\mu\nu}$.
This corresponds to conformal invariance and all of the relevant
equations should display this invariance property. Let us remind
ourselves that $g^{1/2}\rightarrow\Omega^{n/2}g^{1/2}$. Therefore,
conformal invariance is guaranteed if $L_n\rightarrow\Omega^{-n/2}L_n$.
Then $\gamma^{\mu\nu}\rightarrow\Omega^{-n/2}\gamma^{\mu\nu}$. The
relation (\ref{44}) can be rewritten in a way displaying this property
explicitly as

\begin{equation}
\nabla_\lambda\gamma^{\mu\nu}-{1\over n}\,\left(\gamma_{\sigma\tau}\,\nabla_\lambda\gamma^{
\sigma\tau}\right)\,\gamma^{\mu\nu}=0\,.
\label{47}
\end{equation}

With all the ingredients above we can write $\gamma^{\mu\nu}$ as

\begin{equation}
\gamma^{\mu\nu}=\left({{\bigl<R\bigr>}\over{n\Lambda}}\right)^{n/2-1}\,g^{\mu\nu}\,.
\label{48}
\end{equation}

This construction procedure forces us to introduce, for dimensional
reasons, the constant $\Lambda$ with dimensions ${\rm
dim}[\Lambda]=L^{-2}$; the factor $n$ is there just for computational
convenience.

\end{enumerate}

As a consequence of conformal invariance, equation (\ref{47}) is
traceless with respect to $\gamma_{\mu\nu}$ and, therefore, there is an
arbitrariness in the solution of (\ref{47}). The most general solution
of (\ref{47}), incorporating this arbitrariness, is

\begin{equation}
\nabla_\lambda\gamma_{\mu\nu}=-W_\lambda'\,\gamma_{\mu\nu}\,.
\label{49}
\end{equation}

\noindent This equation can be solved for the connection to obtain

\begin{equation}
{\Gamma^\lambda}_{\mu\nu}=\left\{{}^{\,\lambda}_{\mu\nu}\right\}(\gamma)+{W^\lambda}_{\mu\nu}(\gamma,\,{\bf W}')\,,
\label{50}
\end{equation}

\noindent where $\left\{{}^{\,\lambda}_{\mu\nu}\right\}(\gamma)$ is the
Christoffel symbol (\ref{46}) for the tensor $\gamma_{\mu\nu}$ and

\begin{equation}
{W^\lambda}_{\mu\nu}(\gamma,\,{\bf W}')={1\over2}\,\gamma^{\lambda\rho}\,
\left(W_\mu'\,\gamma_{\nu\rho}+W_\nu'\,\gamma_{\mu\rho}-W_\rho'\,\gamma_{\mu\nu}\right)\,,
\label{51}
\end{equation}

\noindent where $W_\mu'$ is an arbitrary Weyl vector field \cite{We26}.

Let us now return to equation (\ref{48}). This relation can be
rewritten as

\begin{equation}
\gamma_{\mu\nu}=e^\Psi\,g_{\mu\nu}\,,
\label{52}
\end{equation}

\noindent where $e^\Psi$ is a dimensionless function given by

\begin{equation}
e^\Psi=\left({{n\Lambda}\over{\bigl<R\bigr>}}\right)^{n/2-1}\,.
\label{53}
\end{equation}

\noindent In terms of this parametrization, the connection is now given
by

\begin{equation}
{\Gamma^\lambda}_{\mu\nu}=\left\{{}^{\,\lambda}_{\mu\nu}\right\}({\bf g})
+{W^\lambda}_{\mu\nu}({\bf g},\,{\bf W})\,,
\label{54}
\end{equation}

\noindent where $W_\mu=W_\mu'+\partial_\mu\Psi$.

Let us finally note that the resulting theory is conformally invariant.
One way of breaking the conformal invariance is to let $\Psi$ acquire a
specific value. Due to the conformal invariance, we can fix this value
at will, in particular, we can choose $\Psi=0$. In this last case we
obtain an effective theory for low energies where conformal invariance
has been broken. In this case the field equations (\ref{36}) and
(\ref{53}), evaluated for the connection (\ref{54}) with $\Psi=0$ are

\begin{eqnarray}
R_{\mu\nu}({\bf g})-{{(n-2)}\over4}\,(\nabla_\mu W_\nu+\nabla_\nu W_\mu)+{{(n-2)}\over4}\,W_\mu
\,W_\nu&&\nonumber\\
-{1\over n}\,\left[R({\bf g})-{{(n-2)}\over2}\,(\nabla W)+{{(n-2)}\over4}\,W^2\right]\,
g_{\mu\nu}&=&0\,,
\label{55}
\end{eqnarray}

\begin{equation}
R({\bf g})-(n-1)\,(\nabla W)+{{(n-1)(n-2)}\over4}\,W^2=n\,\Lambda\,,
\label{56}
\end{equation}

\noindent where now $\nabla$ is the usual covariant derivative with
respect to the metric $g_{\mu\nu}$. The field equations (\ref{55}) and
(\ref{56}) are written completely in terms of the Riemannian metric
$g_{\mu\nu}$ and therefore can be compared, interpreted and tested in
familiar physical situations.

We can now turn to the conclusions. Firstly, we have shown that almost
all Lagrangians depending on
${\bigl<R\bigr>^\mu}_\nu=g^{\mu\lambda}R_{\lambda\nu}({\bf\Gamma})$, in
a generic and almost arbitrary way, possess the same `universal' field
equations, namely (\ref{11}). In the circumstances described under
point (2), it is furthermore possible to incorporate a preferred
direction. The freedom contained in the choice of the starting
Lagrangian may be used to satisfy the several different requirements
usually needed in order to reach a consistent quantum version of
canonical gravity. For $n=4$ the universal field equations are
(\ref{15}) and they correspond to the usual Einstein field equations
with a necessarily non--null cosmological constant and a Weyl vector
field $W_\mu$. Both of these new properties of the gravitational field
equations have strong observational support. Naturally, the next step
is to look for solutions of equation (\ref{15}) in particularly
interesting physical situations. $\Lambda$ and $W_\mu$ are of
cosmological character and their values must be determined from
astrophysical observations. Let us first analyse the case $W_\mu=0$.
Recent observations \cite{OS95} point to a non--null (even when small)
value of the cosmological constant $\Lambda$. A non--null cosmological
constant does not have a significant role over solar system scales. In
fact, the corresponding spherically symmetric solution is the Kottler
metric. The Schwarzschild solution is naturally recovered for
$\Lambda=0$ and for small $\Lambda$, which is the actual case, there
are no observational effects. However, observational effects may be
expected at the level of galactic dynamics \cite{Ma97}. On the other
hand, there is some evidence that the universe is not completely
isotropic, as assumed in the cosmological principle, and recent
observations point to the existence of a preferred direction in the
Universe \cite{NR97}. That direction may be accounted for by the Weyl
vector field appearing in (\ref{15}). The possibility of accommodating
the observed anisotropy in the field equations (\ref{15}) is currently
under study with the use of Bianchi spaces (the natural anisotropic
extension of FRW spaces). Once the value of $W_\mu$ is determined, we
can look for other observational effects of $W_\mu$. However, and in a
way similar to the case of the cosmological constant, we believe that
the magnitude of the anisotropy is small as to play a significant role
over solar system scales. Therefore, observable effects must be sought
at the level of galactic dynamics.

In conclusion, the `universal field equations' (\ref{15}) can be
considered as a viable generalization of the Einstein field equations
of general relativity. Furthermore, almost any metric--affine
Lagrangian is a good candidate for this purpose and this additional
freedom may serve to accommodate several other requirements.

Let us finally add that the universal field equations (\ref{15}) are
not exclusive to the Lagrangians considered here. They have also been
obtained in \cite{TR98}, in the context of the recently developed
fourth--rank gravity \cite{TRMC}.

\section*{Acknowledgements}

The authors are grateful to G. Rubilar for calling their attention to a
fundamental inconsistency in the original formulation of the work.


\end{document}